\documentclass[12pt]{iopart}

\usepackage{amsfonts}
\usepackage{amssymb}
\usepackage{iopams}
\usepackage{setstack}
\usepackage{graphicx}

\newcommand{\ud}{\mathrm{d}}
\newcommand{\ue}{\mathrm{e}}

\newcommand{\intl}{\int_{-\infty}^{+\infty}}
\newcommand{\sign}{\mathrm{sign}}

\begin{document}

\letter{The short pulse equation and associated constraints}

\author{Theodoros P. Horikis}

\address{Department of Computer Science and Technology, University of
Peloponnese, Tripolis 22100, Greece}

\ead{horikis@uop.gr}

\begin{abstract}
The short pulse equation (SPE) is considered as an initial-boundary value problem.
It is found that the solutions of the SPE must satisfy an integral relation
otherwise the temporal derivative exhibits discontinuities. This integral relation
is not necessary for a solution to exist. An infinite number of such constraints
can be dynamically generated by the evolution equation.
\end{abstract}

\submitto{\JPA}

\maketitle

The standard model for describing propagation of a pulse-shaped complex field
envelope in nonlinear dispersive media is the nonlinear Schr\"{o}dinger (NLS)
equation. In the context of nonlinear optics, the main assumption made when
deriving the NLS equation from Maxwell's equations is that the pulse-width is large
as compared to the period of the carrier frequency. When this assumption is no
longer valid, i.e., for pulse duration of the order of a few cycles of the carrier,
the evolution of such ``short pulses'' is better described by the so-called
short-pulse equation (SPE) \cite{schafer}.

The SPE can be expressed in the following dimensionless form,
\begin{equation}
u_{xt}=u+\frac{1}{6}(u^3)_{xx} \label{spe}
\end{equation}
where subscripts denote partial derivatives. The SPE forms an initial-boundary
problem when accompanied by the initial data
\[
u(0,x)=u_0,
\]
and sufficiently fast decaying boundary conditions $u(t,\pm\infty)=0$. Much like
the NLS equation, the SPE is integrable \cite{sakovich3} and exhibits soliton
solutions in the form of loop-solitons \cite{sakovich1}. However, when it is formed
as an evolution equation certain conditions must apply otherwise, as shown below,
the temporal derivative exhibits discontinuities.

Despite the fact that the equation is integrable via the inverse scattering
transform \cite{victor}, there are certain subtleties that need to be clarified.
Integration of Eq. (\ref{spe}) introduces the operation
\[
\partial_x^{-1}u(t,x)=\int_{-\infty}^x
u(t,x')\;\ud x'
\]
Clearly as $x$ approaches $-\infty$, $\partial_x^{-1}u=0$, consistent with rapidly
decaying data. However, as $x$ approaches $+\infty$, for $u$ and its time and space
derivatives to decay, a constraint seems to be necessary (see the discussion
below), namely
\begin{equation}
\intl u(t,x)\;\ud x =0 \label{spe.const}
\end{equation}
Indeed, writing the SPE in evolution type form we have
\[
u_t=\partial_x^{-1}u+\frac{1}{6}(u^3)_x=\int_{-\infty}^x u\;\ud x' + \frac{1}{6}(u^3)_x
\]
and imposing the boundary condition as $x\rightarrow +\infty$, one results to Eq.
(\ref{spe.const}).

In fact, this constraint induces further constraints obtained by successively
taking the time derivative of the integral and using Eq. (\ref{spe}). For example,
the next constraint is given by
\begin{equation}
\intl \partial_x^{-1}u \; \ud x =0\label{spe.const2}
\end{equation}
However, Eqs. (\ref{spe.const}), (\ref{spe.const2}), along with the rest of the
family of infinite constraints generated as above, are not generically true. One
might surmise that constraints are required at all times for a solution to exist.
However, as discussed below, this is not the case. Extra constraints on the initial
data are not necessary, but the solution suffers from a temporal discontinuity. For
smooth initial data not satisfying Eq. (\ref{spe.const}), $u_t(t,x)$ has at $t=0$
different left and right limits and the rest of the family of constraints cannot be
generated dynamically at that point. The same issues arise in the context of the
Kadomtsev-Petviashvili (KP) equations and were studied in Refs. \cite{mja1,mja2}.

Our analysis starts by taking the Fourier transform (FT) of Eq. (\ref{spe}),
\begin{equation}
i k  \hat{u}_t=\hat{u}-\frac{ k ^2}{6}\widehat{u^3} \label{spe.ft}
\end{equation}
where the FT pair is defined as
\begin{eqnarray*}
\hat{u}(t, k )=\mathcal{F}\left\{ u(t,x) \right\}&=&\intl u(t,x)\,\ue^{i k  x}\;\ud x \\
u(t,x)=\mathcal{F}^{-1}\left\{ \hat{u}(t, k ) \right\}&=&\frac{1}{2\pi}
\intl \hat{u}(t, k )\,\ue^{-i k  x}\;\ud  k
\end{eqnarray*}
Define $\hat{U}=\widehat{u^3}$ and write Eq. (\ref{spe.ft}) in the form of a first
order differential equation in $t$,
%
\begin{eqnarray}
\hat{u}_t-\frac{1}{i k }\hat{u}=\frac{i k }{6}\hat{U}
\label{fourier}
\end{eqnarray}
%
which can be readily integrated with the use of integrating factors to give
\begin{equation}
\hat{u}(t, k )=\hat{u}_0\,\ue^{t/i k }+\frac{i k }{6}\,\ue^{t/i k }
\int_0^t \hat{U}\,\ue^{-\tau/i k }\;\ud\tau \label{sol.ft}
\end{equation}
where $\hat{u}_0=\hat{u}_0( k )=\mathcal{F}\{ u_0(x) \}$, is the FT of the initial
data. Using Eq. (\ref{sol.ft}) we calculate the temporal derivative to be
\begin{equation}
\hat{u}_t(t, k )=\frac{1}{i k }\hat{u}_0\,\ue^{t/i k }+\frac{i k }{6}\hat{U}
+\frac{1}{6}
\int_0^t \hat{U}\,\ue^{(t-\tau)/i k }\;\ud\tau \label{der.ft}
\end{equation}
Clearly, from Eq. (\ref{fourier}), as $ k $ tends to zero, we should demand that so
will $\hat{u}$. This translates to \cite{pelinovsky}
\[
\hat{u}(t,0)=0 \Leftrightarrow \intl u(t,x)\;\ud x =0
\]
However, as $t\rightarrow \pm 0$, we have that $\hat{u}(0,x)=\hat{u}_0$ from Eq.
(\ref{sol.ft}), and
\[
\hat{u}_t=\frac{1}{ ik -\,\sign(t)0}\,\hat{u}_0+\frac{i k }{6}\,\hat{U}
\]
This is because the function $\exp(t/ ik )$ defines a distribution, depending
continuously on $t$, in the Schwartz space of the variable $ k $ \cite{boiti} with
\[
\frac{\partial}{\partial t}\ue^{t/ ik }=\frac{1}{ ik -\,\sign(t)0}\,\ue^{it/ k },
\quad t=0
\]
and
\[
\frac{\partial}{\partial t}\ue^{t/ ik }=\frac{1}{ ik }\,\ue^{t/ ik },
\quad t\ne 0
\]
This suggests that although there is no discontinuity in the solution, there is one
in the derivative. Indeed, taking the inverse FT of Eq. (\ref{der.ft}), at
$t\rightarrow \pm 0$, we have
\[
u_t(t\rightarrow \pm 0,x)=\frac{1}{2\pi}\lim_{t\rightarrow \pm 0} \intl
\frac{1}{i k }\hat{u}_0\, \ue^{t/i k }\,\ue^{-i k  x}\;\ud k
+\frac{1}{6}(u^3)_{x} (t\rightarrow \pm 0,x)
\]
The nonlinear term is straightforward to handle so we focus on the linear part,
\begin{eqnarray}
I(x)&=& \frac{1}{2\pi}\lim_{t\rightarrow \pm 0} \intl
\frac{1}{i k }\hat{u}_0\, \ue^{t/i k }\,\ue^{-i k  x}\;\ud k  \nonumber\\
&=&\frac{1}{2\pi}\lim_{t\rightarrow \pm 0} \intl
\frac{1}{i k }\left[\hat{u}_0\ue^{-i k  x} +\hat{u}_0(0) - \hat{u}_0(0) \right]
\ue^{t/i k }\;\ud k  \nonumber\\
&=& \frac{1}{2\pi}\lim_{t\rightarrow \pm 0} \intl
\frac{1}{i k }\left[\hat{u}_0\ue^{-i k  x} - \hat{u}_0(0) \right]
\ue^{t/i k }\;\ud k  \nonumber\\ &+& \frac{1}{2\pi}\lim_{t\rightarrow \pm 0} \intl
\frac{1}{i k }\,\hat{u}_0(0)
\,\ue^{t/i k }\;\ud k \label{ix}
\end{eqnarray}
Using the property
\[
\intl \frac{1}{i k }\,\ue^{t/i k }\;\ud k =-\pi\,\sign(t)
\]
the second integral of Eq. (\ref{ix}) is reduced to $-
\hat{u}_0(0)\pi\,\sign(t)/2$. Furthermore, we write
\[
\hat{u}_0(0)=\intl \delta( k )\,\hat{u}_0( k )\,\ue^{-i k  x}\;\ud k
\]
so that finally
\begin{eqnarray*}
I(x)&=&\frac{1}{2\pi}\intl \left[ \mathrm{P}\left(\frac{1}{i k }\right)
-\pi\,\sign(t)\delta( k )\right]\hat{u}_0( k )\,\ue^{-i k  x}\;\ud k
\\ &=& \frac{1}{2\pi}\intl \frac{\hat{u}_0( k )}{i k -0\,\sign(t)}\,
\ue^{-i k  x}\;\ud k \\ &=& \int_{\sign(t)\infty}^{x}u_0(x')\;\ud x'
\end{eqnarray*}
where P denotes principal value. Thus, at $t=0$, Eq. (\ref{der.ft}) translates into
physical space as
\[
u_t =\int_{\sign(t)\infty}^{x}u_0(x')\;\ud x' +\frac{1}{6}(u_0)_{x}
\]
As also mentioned in Refs. \cite{pelinovsky,mja2}, the operator
\[
\partial_x^{-1}=\int_{\sign(t)\infty}^{x} \;\ud x'
\]
and its relative average
\[
\partial_x^{-1}=\frac{1}{2}\left( \int_{-\infty}^{x} \;\ud x' + \int_{x}^{\infty} \;\ud x'\right)
\]
are equivalent, meaning that one can choose either one of them. If $t\ne 0$ we have
that $\intl u\;\ud x=0$, hence both choices are valid. At $t=0$ there is a
discontinuity in the temporal derivative.

For the evolution of the SPE, Eq. (\ref{spe.const}) is not preserved in time and as
such leads to the infinite number of further constraints. Indeed, if Eq.
(\ref{spe.const}) holds then an infinite number of constraints, dynamically
generated using the SPE, hold during the evolution. Within the physical framework
of the SPE these constraints are neither ``natural" nor necessary. Solutions of the
SPE can exist without satisfying this condition, the most prominent example being
the loop-soliton \cite{sakovich1}. This solution, however, in addition to the
possible temporal discontinuities, suffers from  discontinuities in its spatial
derivatives, $u_x(t,x)$, and extra care may be needed when the above formalism is
applied.

We conclude with a note on the so-called regularized SPE (RSPE) model, recently
derived in Ref. \cite{costanzino}. The latter has been derived by including a
regularization term, based on the next term in the expansion of the dielectric's
susceptibility. In that case, the pulses (of the real component of the electric
field) are described by:
\[
u_{xt}=u+\frac{1}{6}(u^3)_{xx}+\beta u_{xxxx}
\]
where $\beta$ is a small parameter. Without the regularization term, $\beta
u_{xxxx}$, i.e., in the case of the SPE --cf. Eq. (\ref{spe})--, traveling pulses
in the class of piecewise smooth functions with one discontinuity do not exist.
However, when the regularization term is added, and for a particular parameter
regime, the RSPE supports smooth traveling waves which have structure similar to
solitary waves of the modified KdV equation \cite{costanzino}. The regularization
term does not alter the analysis for the SPE. Indeed, in the Fourier domain the
term is written as $\mathcal{F}\left\{ \beta u_{xxxx}\right\}=\beta k ^4\hat{u}$
and thus when dividing with $ik$ from the left-hand-side the resulting power of $ k
$ is continuous at $k=0$. The linear part of the RSPE, much like the linear part of
the KP-I equation \cite{mja1,boiti}, deserves more study and the analysis will be
presented in a future communication.
\section*{Acknowledgments}
I wish to thank Mark J. Ablowitz for bringing the KP analysis to my attention and Barbara
Prinari, Dimitri J. Frantzeskakis and
Panayotis G. Kevrekidis for many useful discussions.

\end{document}